\documentclass[prl,twocolumn,showpacs,superscriptaddress]{revtex4}

\usepackage{graphicx,epsfig,color}

\usepackage{mathrsfs} 
\usepackage{amsmath}
\usepackage{amssymb}
\usepackage[utf8]{inputenc}  
\usepackage[english,francais]{babel}

\usepackage{url,hyperref}  

\usepackage{enumitem}

  \definecolor{blue}{rgb}{0,0,1}
  \definecolor{green}{rgb}{0,.6,0}
  \definecolor{red}{rgb}{1,0,0}
  \definecolor{vio}{rgb}{1,0,1}
  \definecolor{uv}{rgb}{0.5,0,0.5}
  \definecolor{ama}{rgb}{0.3,0.3,0.3}
\definecolor{M_Beige}         {rgb}{0.96 , 0.96 , 0.86}

\definecolor{M_Brown}         {rgb}{0.65 , 0.16 , 0.16}

\definecolor{M_Gold}          {rgb}{1.00 , 0.84 , 0.00}

\definecolor{M_LemonChiffon}  {rgb}{1.00 , 0.98 , 0.80}

\definecolor{M_Orange}        {rgb}{1.00 , 0.60 , 0.00}

\definecolor{M_Pink}          {rgb}{0.80 , 0.55 , 0.60}

\definecolor{M_Violet}          {rgb}{0.83 , 0.21 , 0.93}

\definecolor{M_Green}          {rgb}{0.2 , 0.6 , 0.2}

\definecolor{M_Gray}          {rgb}{0.7 , 0.7 , 0.7}

\definecolor{M_BluPal}          {rgb}{0.7 , 0.7 , 0.9}

\def\XXint#1#2#3{{\setbox0=\hbox{$#1{#2#3}{\int}$}
\vcenter{\hbox{$#2#3$}}\kern-.5\wd0}}

\renewcommand{\geq}{\geqslant}

\newcommand{\EXP}[1]{\mathrm{e}^{#1}}         

\def\eqlaw{\stackrel{\mbox{\tiny (law)}}{=}}     
\newcommand{\ket}[1]{|\kern.3ex#1\kern.3ex\rangle}
\newcommand{\bra}[1]{\langle\kern.3ex #1 \kern.3ex|}

\newcommand{\braket}[2]{\langle\kern.3ex #1 \kern.3ex|\kern.3ex #2 \kern.3ex \rangle}
\newcommand{\braketYves}[2]{\left(\kern.3ex #1 \kern.3ex|\kern.3ex #2 \kern.3ex \right)}

\newcommand{\mean}[1]{\left\langle #1 \right\rangle} 
\newcommand{\smean}[1]{\langle #1 \rangle} 

\newcommand{\heaviside}{\mathop{\theta_\mathrm{H}}\nolimits}  

\def\D{{\rm d}}                  

\newcommand\antiddots{\mathinner{\mkern2mu\raise1pt\hbox{.}\mkern2mu
    \newline \raise4pt\hbox{.}\mkern2mu\raise7pt\hbox{.}\mkern1mu}}

\def\mass{m}
\def\Filoche{ {\mbox{\tiny{ADJMF}}} }

\def\IDoS{\mathscr{N}}

\begin{document}

\selectlanguage{english}

\title{Comment on \og Effective Confining Potential of Quantum States in Disordered Media \fg{}  }

\author{Alain Comtet}
\affiliation{LPTMS, CNRS, Univ.~Paris-Sud, Universit\'e Paris-Saclay, F-91405 Orsay, France}
\author{Christophe Texier}
\affiliation{LPTMS, CNRS, Univ.~Paris-Sud, Universit\'e Paris-Saclay, F-91405 Orsay, France}

\date{\today}




\begin{center}
\textbf{\large Comment on \og Effective Confining Potential of Quantum States in Disordered Media \fg{}} \href{https://journals.aps.org/prl/abstract/10.1103/PhysRevLett.116.056602}{[Phys. Rev. Lett. \textbf{116}, 056602 (2016)]}
\end{center}

In the Letter \cite{ArnDavJerMayFil16}, the inverse of the landscape function $u(x)$ introduced in Ref.~\cite{FilMay12} was shown to play the role of an effective potential. This leads to the following estimation of the integrated density of states (IDoS), in~1D,
\begin{equation}
  \label{eq:IDoS-FM}
  \IDoS_\Filoche(E) = \frac{1}{\pi}\int_{u(x)>1/E}\D x\, \sqrt{E-1/u(x)}
  \:.
\end{equation}
We consider here two disordered models for which we obtain the distribution of  $u(x)$  and argue that the precise spectral singularities are not reproduced by Eq.~\eqref{eq:IDoS-FM}. 

\vspace{0.125cm}

\noindent\textit{Pieces model.---} 
We consider the Schr\"odinger Hamiltonian $H=-\D^2/\D x^2 + \sum_n v_n\,\delta(x-x_n)$, where the positions of the $\delta$ potentials are independently and uniformly distributed on $[0,L]$ with mean density $\rho$.
The landscape function, which solves $Hu(x)=1$, is thus parabolic on each free interval.
In the limit $v_n\to+\infty$ (\og pieces model \fg{}), intervals between impurities decouple 
and IDoS per unit length is
$N(E)=\lim_{L\to\infty}(1/L)\IDoS(E)=\rho/\big[\EXP{\pi\rho/\sqrt{E}}-1\big]$ \cite{LutSy73}. 
We compare it with \eqref{eq:IDoS-FM}. 
Assuming now ordered positions, $x_1<x_2<\cdots$, we have $u(x)=(1/2)(x-x_{n-1})(x_n-x)$ for $x\in[x_{n-1},x_n]$.
We first study its distribution $P(u)=\mean{\delta(u-u(x)}$. 
The disorder average can be replaced by a spatial average, 
$P(u)=\rho^2\int_0^\infty\D\ell\,\EXP{-\rho\ell}\int_0^\ell\D x\,\delta(u-x(\ell-x)/2)$, leading to
\begin{equation}
  P(u) 
  = 4\rho^2\,K_0(\rho\sqrt{8u})
  \:,
\end{equation}
where $K_\nu(z)$ is the MacDonald function. 
Denoting by $\heaviside(x)$ the Heaviside function,
we can now deduce the estimate 
$N_\Filoche(E)=(1/\pi)\smean{\sqrt{E-1/u}\,\heaviside(E-1/u)}$~:
\begin{align}
  N_\Filoche(k^2) 
  = \frac{k}{\pi} \int_\xi^\infty
    \hspace{-0.25cm}\D t\, 
  \sqrt{t^2-\xi^2}\,K_0(t)
  \hspace{0.2cm}\mbox{for }
  \xi=\frac{\rho\sqrt{8}}{k}
\end{align}
For $k=\sqrt{E}\gg\rho$, we get $N_\Filoche(k^2)\simeq k/\pi$, as it should.
For low energy, $k\ll\rho$, one gets 
$N_\Filoche(k^2)\simeq (k/2)\,\exp\{-\sqrt{8}\rho/k\}$, which is a rather poor approximation of the Lifshitz tail $N(k^2)\simeq \rho\,\exp\{-\pi\rho/k\}$~: the coefficient in the exponential is underestimated and the preexponential function incorrect, thus overestimating the IDoS  by an exponential factor.

\vspace{0.125cm}

\noindent\textit{Supersymmetric quantum mechanics.---}
We consider the Hamiltonian \cite{ComTex98} 
$H=Q^\dagger Q$, 
where $Q=-\partial_x + \mass(x)$. 
The analysis is more simple for boundary conditions 
$\psi(0)=0$ \& $Q\psi(L)=0$, leading to the 
Green's function 
$
  G(x,y) =\bra{x}H^{-1}\ket{y} =  \psi_0(x)\psi_0(y) \int_0^{\mathrm{min}(x,y)}\D z\,\psi_0(z)^{-2}
$,
where  $\psi_0(x)=\exp\big\{\int_0^x\D t\,\mass(t)\big\}$.
We study $u(x)=\int_0^L\D y\,G(x,y)$ when $\mass(x)$ is a Gaussian white noise with $\mean{\mass(x)}=\mu\,g$ and $\mean{\mass(x)\mass(x')}_c=g\,\delta(x-x')$, thus $B(x)=\int_0^x\D t\,\mass(t)$ is a Brownian motion (BM) with drift $\mu$
(in Ref.~\cite{ComDesMon95}, the more regular case with $\mass(x)$ being a random telegraph process was considered, leading to the same low energy properties).
We have
\begin{align}
  \label{eq:NiceUforDN}
   u(x) &= 
   \EXP{B(x)}
   \bigg\{
     \int_0^x\D y\,\EXP{B(y)}\int_0^y\D z\,\EXP{-2B(z)}
     \nonumber\\
     +&\int_0^x\D y\,\EXP{-2B(y)}\int_x^L\D z\,\EXP{B(z)}   
   \bigg\}\equiv u_<(x) + u_>(x)
\end{align}
The cases $\mu\geq0$ and $\mu<0$ are very different:
numerical simulations show that the first moments of $\ln u(x)$ grow with $x$ for $\mu\geq0$ (in particular $\mean{\ln u(x)}\simeq\mu\,g x+\mathrm{cst}$ for $\mu>0$), while they remain uniform (apart near boundaries) for  $\mu<0$. 
%
%
We first discuss the term $u_>(x)=\int_x^L\D y\,G(x,y)$ of \eqref{eq:NiceUforDN}, which is the product of two \textit{independent} exponential functionals of the BM 
$u_>(x) \eqlaw ({4}/{g^2})\,   Z_{gx}^{(-\mu)}\, \widetilde{Z}_{g(L-x)/4}^{(-2\mu)}$, where 
$Z_L^{(\mu)} = \int_0^L \D t\,\EXP{-2\mu t+2W(t)}$, $W(t)$ being a Wiener process (a normalized BM with no drift).
The $n$th moment of $Z_L^{(\mu)}$ is $\sim\EXP{2n(n-\mu)L}$ \cite{MonCom94}, 
thus $\mean{ u_>(x)^n  }\sim \exp\big\{\frac12n^2g(L+3x)+n\,\mu\,g(L+x)\big\}$, which suggests a log-normal tail.
For $\mu\geq0$, there is no limit law and $u_>(x)$ grows exponentially, hence the bound of the landscape approach is useless.
For $\mu<0$, $1/Z_\infty^{(-\mu)}$ is distributed by a Gamma law 
 \cite{MonCom94} and we get the exact distribution of $u_>(x)$ for $x$ \& $L-x\to\infty$~: 
\begin{equation}
  P_>(u) = \frac{2g^{-3|\mu|}u^{-1-3|\mu|/2}}{\Gamma(|\mu|)\Gamma(2|\mu|)}\,
  K_{|\mu|}\!\left(\frac{2}{g\sqrt{u}}\right) 
  \underset{u\to\infty}{\sim} u^{-1-|\mu|}  
  \,.
\end{equation}
$u_<(x)=\int_0^x\D y\,G(x,y)$ should have the same statistical properties, as confirmed numerically. 
Although $u_>(x)$ and $u_<(x)$ are correlated, the distribution of their sum is expected to present the same power law tail $P(u)\sim u^{-1-|\mu|}$, what we checked numerically. 

We now apply \eqref{eq:IDoS-FM}~:
for $\mu\geq0$, $u(x)$ has not limit law when $x$ \& $L-x\to\infty$ and the distribution of $W=1/u(x)$ converges to $\delta(W)$, hence $N_\Filoche(E)= \sqrt{E}/\pi$.
For $\mu<0$, we get $N_\Filoche(E)= (1/\pi)\int_{1/E}^\infty\D u\,P(u)\,\sqrt{E-1/u}\sim E^{|\mu|+1/2}$ for $E\to0$,
while the exact IDoS behaves as $N(E)\sim E^{|\mu|}$ \cite{BouComGeoLeD90}. 
Hence, Eq.~\eqref{eq:IDoS-FM} predicts a power law with an incorrect exponent, i.e. underestimates the IDoS.

For boundary conditions $\psi(0)=\psi(L)=0$, we have also obtained $P(u)\sim u^{-1-|\mu|}$ and $N_\Filoche(E)\sim E^{|\mu|+1/2}$, independently of the sign of $\mu$ in this case.

\bigskip

 Alain Comtet and Christophe Texier

\hfill
\begin{minipage}[t]{0.4\textwidth}
{\small
LPTMS, 
\\
Universit\'e Paris-Saclay, CNRS, 
\\
F-91405 Orsay, France
}
\end{minipage}


\vspace{0.5cm}

\begin{figure}[!ht]
\centering
\includegraphics[width=0.475\textwidth]{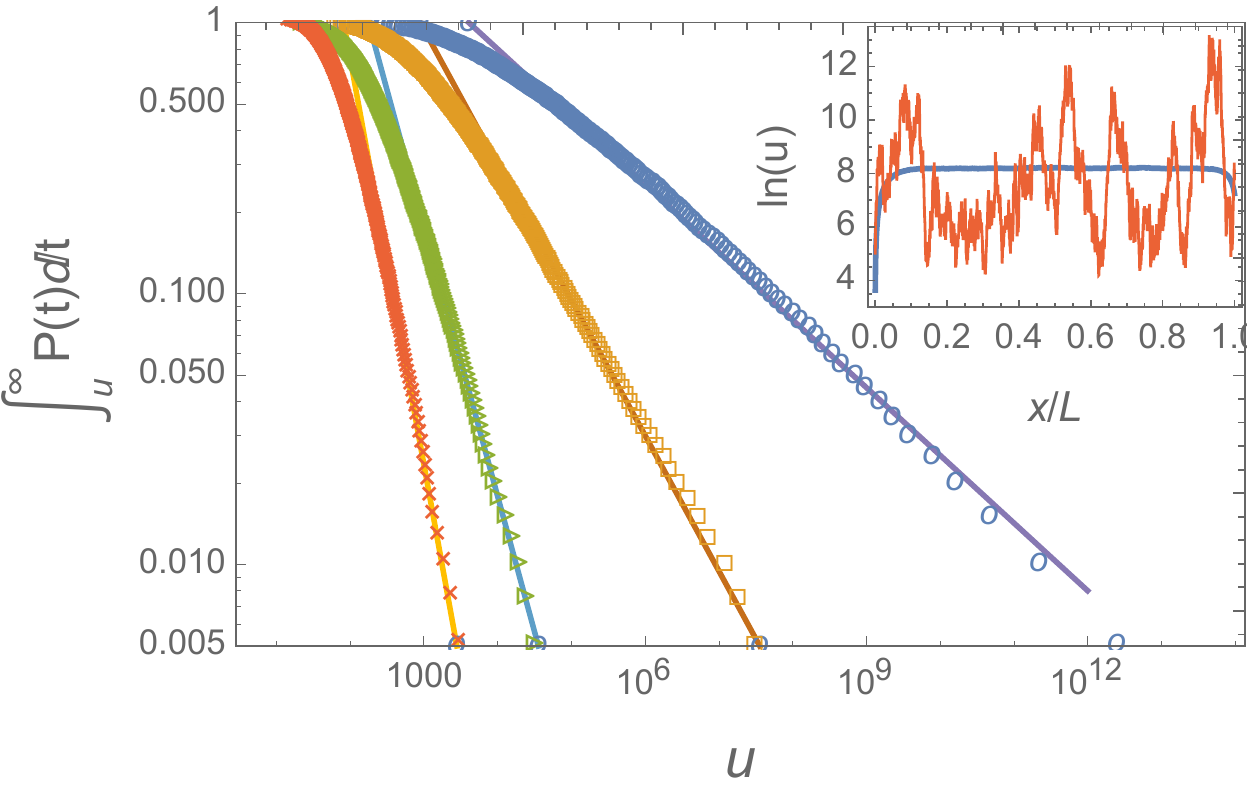}
\caption{{\it Cumulative distribution $\int_u^\infty\D u'\,P(u')$ of the landscape function for drifts $\mu=-0.25$, $\mu=-0.5$, $\mu=-1$ and $\mu=-1.5$ ($gL=100$; $n_s=10^4$ disorder realizations). Straight lines correspond to the power law $u^{-|\mu|}$.}
Inset~: 
\textit{$\ln u(x)$ for $gL=200$ and $\mu=-0.5$ (red line), and $\smean{\ln u(x)}$ after averaging over $n_s=50\,000$ realizations (blue line).}}
\label{fig:PdeUsusy}
\end{figure}

\noindent\textbf{Appendix (arXiv version)~: numerics.---}
The form \eqref{eq:NiceUforDN} is appropriate for numerical simultation.
In the inset of Fig.~\ref{fig:PdeUsusy}, we plot the result of a numerical simulation for $\mu<0$ for one realization of the disorder. We also plot $\mean{\ln u(x)}$, which is uniform in the bulk (while for $\mu>0$, it grows linearly, $\mean{\ln u(x)}\simeq\mu\,g x+\mathrm{cst}$).
Then we study its distribution and check the limiting behaviour $P(u)\sim u^{-1-|\mu|}$.
 
\end{document}